\documentclass[dvipdfmx,12pt]{elsarticle}
\usepackage{hyperref}
\usepackage[switch,columnwise]{lineno}
\usepackage{amssymb}
\makeatletter
\def\ps@pprintTitle{%
  \let\@oddhead\@empty
  \let\@evenhead\@empty
  \let\@oddfoot\@empty
  \let\@evenfoot\@oddfoot
}
\makeatother

\usepackage{float}

\begin{document}

\begin{frontmatter}

%% Title, authors and addresses

%% use the tnoteref command within \title for footnotes;
%% use the tnotetext command for theassociated footnote;
%% use the fnref command within \author or \address for footnotes;
%% use the fntext command for theassociated footnote;
%% use the corref command within \author for corresponding author footnotes;
%% use the cortext command for theassociated footnote;
%% use the ead command for the email address,
%% and the form \ead[url] for the home page:
%% \title{Title\tnoteref{label1}}
%% \tnotetext[label1]{}
%% \author{Name\corref{cor1}\fnref{label2}}
%% \ead{email address}
%% \ead[url]{home page}
%% \fntext[label2]{}
%% \cortext[cor1]{}
%% \affiliation{organization={},
%%             addressline={},
%%             city={},
%%             postcode={},
%%             state={},
%%             country={}}
%% \fntext[label3]{}

\title{Lowering the Energy Threshold using a Plastic Scintillator and Radiation-Damaged SiPMs
}

%% use optional labels to link authors explicitly to addresses:
%% \author[label1,label2]{}
%% \affiliation[label1]{organization={},
%%             addressline={},
%%             city={},
%%             postcode={},
%%             state={},
%%             country={}}
%%
%% \affiliation[label2]{organization={},
%%             addressline={},
%%             city={},
%%             postcode={},
%%             state={},
%%             country={}}

\author[inst1]{Teruaki Morishita}

\affiliation[inst1]{organization={Department of Physics, Graduate School of Advanced Science and Engineering, Hiroshima University},%Department and Organization
            addressline={1-3-1 Kagamiyama}, 
            city={Higashi-Hiroshima, Hiroshima},
            postcode={739-8526}, 
            country={Japan}}

\author[inst1]{Yasushi Fukazawa}
\author[inst1]{Hiromitsu Takahashi}
\author[inst1]{Taishu Kayanoki}
\author[inst1]{Ryota Niwa}
\author[inst1]{Masaki Hashizume}

\begin{abstract}
%% Text of abstract

The radiation damage to a silicon photomultiplier (SiPM) set on a satellite orbit increases energy threshold for scintillator detectors. We confirmed that 1 krad of radiation increases the energy threshold by approximately a factor of 10, which is worst for our system. Using one or two SiPMs damaged by proton irradiation and a plastic scintillator, we performed the following three experiments in our attempt to lower the energy threshold of radiation-damaged SiPMs to the greatest extent: (1) measurements using a current waveform amplifier rather than a charge-sensitive amplifier, (2) coincidence measurements with two radiation-damaged SiPMs attached to one scintillator and summing up their signals, and (3) measurements at a low temperature. Our findings confirmed that the use of a current waveform amplifier, as opposed to a charge-sensitive amplifier and a shaping amplifier, could lower the energy threshold to approximately 65$\%$ (from 198 keV to 128 keV). Furthermore, if we set the coincidence width appropriately and sum up the signals of the two SiPMs in the coincidence measurement, the energy threshold could be lowered to approximately 70$\%$ (from 132 keV to 93 keV) with little loss of the acquired signal, compared to that of use of only one scintillator. Finally, if we perform our measurements at a temperature of -20 $^\circ$C, we could lower the energy threshold to approximately 34$\%$ (from 128 keV to 43 keV) compared to that of at 20 $^\circ$C. Accordingly, we conclude that the energy threshold can be lowered to approximately 15$\%$ by using a combination of these three methods.
\end{abstract}

%%Graphical abstract
%\begin{graphicalabstract}
%\includegraphics{grabs}
%\end{graphicalabstract}

%%Research highlights
\begin{highlights}
\item Radiation damage increases the energy thresholds of SiPMs.
\item Using a current waveform amplifier can lower the energy threshold.
\item Coincidence measurement with two SiPMs and a scintillator can lower the energy threshold.
\item Measurement at low temperature can lower the energy threshold.
\end{highlights}

\begin{keyword}
%% keywords here, in the form: keyword \sep keyword
Gamma-ray Detector; Plastic Scintillator; Silicon Photomultiplier (SiPM, MPPC); Coincidence; Current Waveform Amplifier; Low Temperature
%% PACS codes here, in the form: \PACS code \sep code
%\PACS 0000 \sep 1111
%% MSC codes here, in the form: \MSC code \sep code
%% or \MSC[2008] code \sep code (2000 is the default)
%\MSC 0000 \sep 1111
\end{keyword}

\end{frontmatter}

%%\linenumbers

%% main text
\section{Introduction}\label{setu3}
In soft gamma-ray observations, the entire detector is generally surrounded by an active shield that can effectively reduce potential background events using an anti-coincidence technique. Recent soft gamma-ray observation satellites often use $\mathrm{Be_{4}Ge_{3}O_{12}}$ (BGO) inorganic scintillators as an active shield, that facilitate high-sensitivity observations due to their high blocking capacity against gamma rays and charged particles\cite{bgo}. However, BGO has the disadvantages of heavy mass and slow decay time, which is in the range of several hundred ns. Organic plastic scintillators, which are composed of C and H, are approximately one-seventh the density of BGO, low-cost, and exhibit a fast decay time of several ns. Additionally, the light yield of plastic scintillator is similar to that of BGO. Ce:$\mathrm{Gd_{3}Al_{2}Ga_{3}O_{12}}$ (GAGG) exhibits a large light yield, however BGO and GAGG are heavy and thus not suitable for CubeSats. For example, in the case of 1U CubeSat, if we use BGO plates with dimensions of $10 cm\times10 cm\times3 cm$, five plates are required to cover; hence, the weight of CubeSat becomes 10 kg. This weight will limit the resources allocated for other subsystems onboard the satellite. A lighter scintillator is better for satellite attitude control or orbit control. As a result, detectors combining a plastic scintillator and a silicon photomultiplier (SiPM) are suitable for small satellites such as the CubeSats. However, the problem with this setup is that when SiPMs are in the satellite orbit, dark current is bound to increase drastically due to the radiation damage caused by cosmic rays, especially by protons\cite{rad}\cite{camelot}. Therefore, the present study attempts to address this issue by lowering the energy threshold of a detector consisting of a plastic scintillator and radiation-damaged SiPMs. In this paper, we define the energy threshold as the limit below which cannot be detected.

First, the plastic scintillator has a short decay time of several ns and noise has various time scales; hence, the signal to noise ratio of the pulse height is larger for the SiPM raw output than that after the integration of what by a charge-sensitive amplifier. Therefore, we consider that the energy threshold could be lowered by the amplification of a current waveform amplifier rather that of a charge-sensitive amplifier. Second, since noise is a randomly occurring phenomenon, we investigated whether it was possible to cut off as much detected noise as possible and lower the energy threshold by the acquisition of only coincidence events with two SiPMs. This was accomplished by adjusting the coincidence window, and summing up the signals of the two SiPMs. Third, we considered reducing the dark current and lowering the energy threshold by measuring at lower temperature because the thermal excitation of carriers in the semiconductor detector is suppressed in that temperature range\cite{hirade}. Section $\ref{setup}$ introduces our experimental setups, and our results are presented in sections $\ref{setu}$, $\ref{setu2}$, and $\ref{low2}$. Finally a summary of our findings is provided in section $\ref{summary}$.

\section{Experimental Settings}\label{setup}
\subsection{Measurement Setup}
\begin{figure}[H]
 \centering
 \includegraphics[scale=0.6]{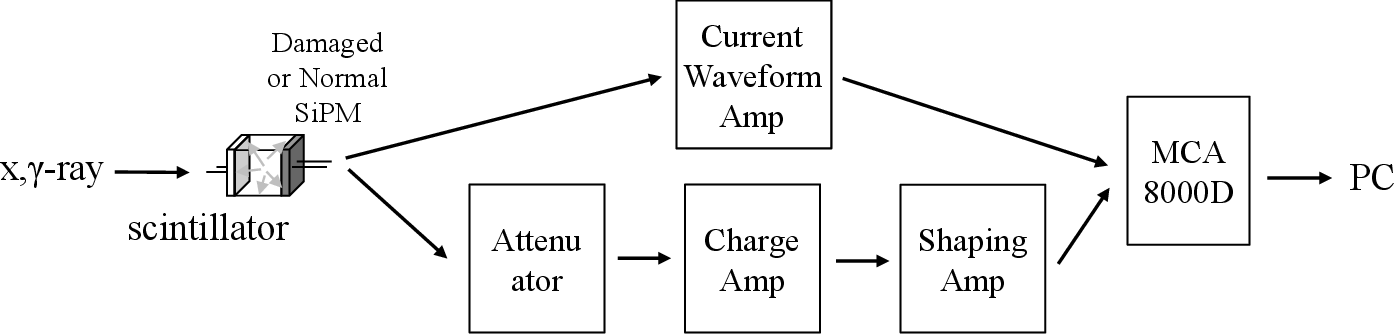}
 \caption{Experimental setup of a single SiPM with MCA 8000D}
 \label{2_4}
\end{figure}

\begin{figure}[H]
 \centering
 \includegraphics[scale=0.8]{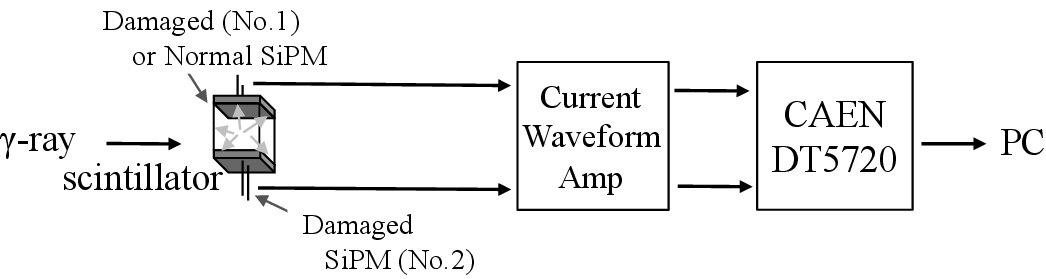}
 \caption{Experimental setup for detecting coincidence events between two SiPMs with CAEN DT5720}
 \label{2_1}
\end{figure}

\begin{figure}[H]
 \centering
 \includegraphics[scale=0.9]{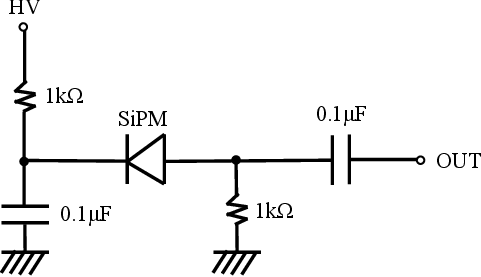}
 \caption{Bias and readout circuit for the SiPMs employed in the present study}
 \label{2_3}
\end{figure}

The following experimental setups were used in the present study. As shown in Figure $\ref{2_4}$ and $\ref{2_1}$, one or two SiPMs with an area of 6 $mm^{2}$ (S13360-6050CS, MPPC provided by Hamamatsu Photonics K.K.\cite{SiPM}) are attached to one or both sides of a 1 cm cubic plastic scintillator (EJ-200\cite{ej200}) with optical grease (TSK5353, Momentive Performance Materials Inc.). Consequently, the sides of the scintillator were wrapped (5-6 laps) with a reflector (P.T.F.E. THREAD SEAL TAPE manufactured by Azwan Inc.). Based on the setup used in a previous study, Figure $\ref{2_3}$ demonstrates the circuit of bias voltage supply and readout of SiPMs\cite{takahashi},\cite{torigoe}. We applied a bias voltage to SiPMs using source meters (2410, 2470 Keithley) in the thermal chamber which is used to maintain temperature at 20 $^\circ$C or -20 $^\circ$C.

Figure $\ref{2_4}$ demonstrates the setup of one of the SiPM readouts. The SiPM signal is fed into the current waveform amplifier (NO8-11 OCTAL PULSE AMPLIFIER, LeCroy) or the charge amplifier (FAST QUAD PREAMPLIFIER/MODEL 5028, CLEAR-PULSE). For the latter, the amplified signal is fed into the shaping amplifier Type 4077-4 (shaping time of 50 ns, CLEAR-PULSE), and an attenuator to prevent saturation. Then, an analog digital converter (MCA 8000D, Amptek Inc.) was used to obtain the pulse height\cite{mca}. We used this setup for two experiments other than the coincidence measurement. Figure $\ref{2_1}$ shows the coincidence measurement setup. Here, two SiPM signals are fed into the current waveform amplifier with an amplification factor of 10, and both the coincidence events and digitization were obtained by the CAEN DT5720 board (CAEN SpA\cite{caen}), a digitizer with a 12-bit ADC resolution and a 250 MS/s sampling rate (1 S/4 ns), which only acquires triggered signals above a predefined trigger threshold. The radiation-damaged SiPMs used in this study were irradiated with 200 MeV protons at 1 krad ($1.71\times10^{10} protons/cm^{2}$) or $6\times10^{9} neutrons/cm^{2}$ in 1 MeV neutron equivalent. The proton irradiation tests were performed in the Wakasa Wan Energy Research Center (Japan) \cite{wakasa}. We used 200 MeV here considering South-Atlantic Anomaly Passages in the low-Earth orbit \cite{hirade}, and the 10 Gy (1 krad) dose corresponds to approximately 1-year operation of satellite without any shields at an altitude of $\sim$500-600 km \cite{camelot}.

\subsection{SiPMs and Bias Voltage}
In this paper, we used two radiation-damaged SiPMs and one normal SiPM. Since $V_{op}=V_{br}+3 V$ where $V_{op}$ is the operation voltage and $V_{br}$ is the breakdown voltage, we set the bias voltage of the first SiPM at 54.75 V (dark current of $\sim$0.3 mA), as shown in Table $\ref{3_1}$. In addition, we adjusted the bias voltage for the second SiPM to ensure that its gain would become almost the same as that of the first SiPM, and then set its bias voltage at 55.4 V (dark current of $\sim$0.3 mA). As shown in Figure \cite{mca}, we used a 1 k$\Omega$ resistor in the circuit, and the voltage provided to SiPMs was reduced by 0.3 V due to the dark current of the radiation-damaged SiPM. It should be noted that although a lower voltage is a better choice for reducing dark current in radiation-damaged SiPMs, the signal generated at this voltage is too small to detect. Accordingly, we set the voltage in Table $\ref{3_1}$.

\begin{table}[htbp]
\centering
\begin{tabular}{c|c c} \hline
S13360-6050CS & $V_{\mathrm{br}}$ (V) & $V_{\mathrm{op}}$ (V) \\ \hline
First (Damaged) & 51.73 & 54.75 \\
Second (Damaged) & 52.68 & 55.40 \\
Third (Normal) & 51.2 & 54.2 \\ \hline
\end{tabular}
\caption{The SiPMs used in the present study}
\label{3_1}
\end{table}

\subsection{Radiation Isotopes}
As shown in Table $\ref{3_6}$, we used radioactive isotopes for energy calibration by utilizing the energies of the photo absorption peak ($\mathrm{E}_{abs}$) and Compton edge ($\mathrm{E}_{edge}$).

\begin{table}[htbp]
\centering
\begin{tabular}{c|c c} \hline
RI & $E_{\mathrm{abs}}$ (keV) & $E_{\mathrm{edge}}$ (keV) \\ \hline
$^{137}\mathrm{Cs}$ & 662 & 477 \\
$^{22}\mathrm{Na}$ & 511 & 341 \\ 
$^{133}\mathrm{Ba}$ & 356 & 196 \\
$^{57}\mathrm{Co}$ & 122 &  \\ 
$^{109}\mathrm{Cd}$ & 88 & \\
$^{241}\mathrm{Am}$ & 59.5 & \\  
$^{109}\mathrm{Cd}$ & 22.2 & \\ \hline
\end{tabular}
\caption{X-ray and gamma-ray energies of the photo absorption peak and Compton edge for the radioisotopes used in the present study for energy calibration}
\label{3_6}
\end{table}

\subsection{CAEN DT5720 Setup}
\begin{figure}[H]
 \centering
% \begin{subfigure}
   \centering
   \includegraphics[scale=0.8]{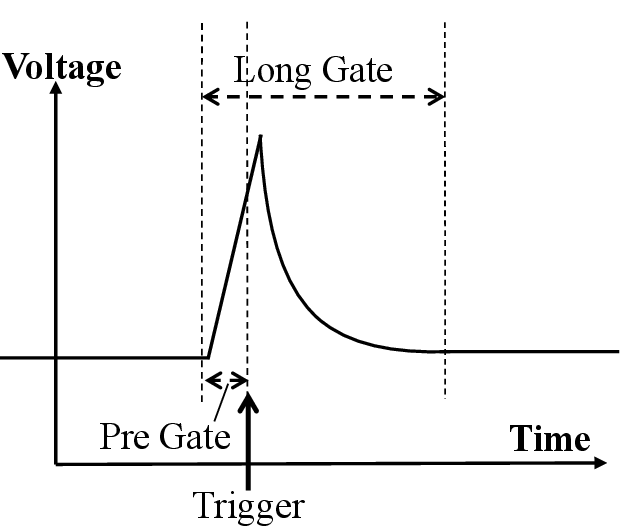}
   \label{3_10}
% \end{subfigure}\vfill
%  \begin{subfigure}
   \centering
   \includegraphics[scale=0.7]{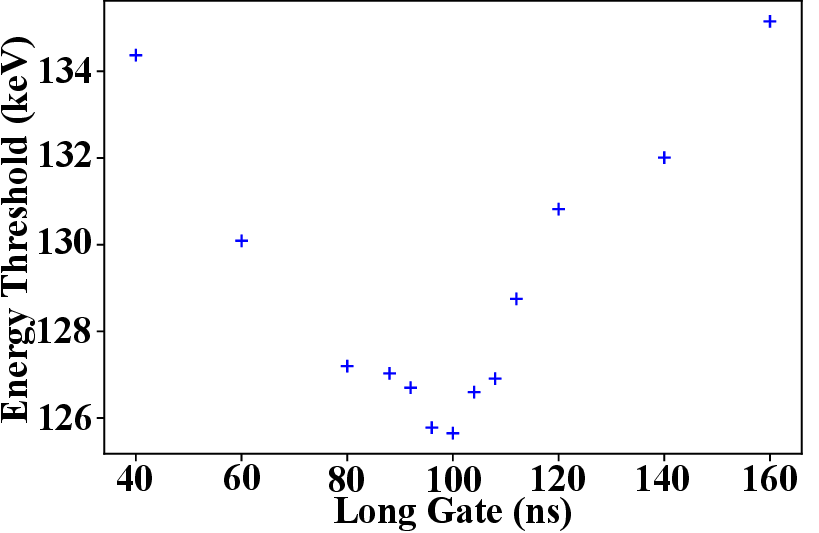}
   \label{3_4}
% \end{subfigure}\vfill
 \caption{(Top) Schematic view of the waveform and CAEN DT5720 gates (based on \cite{caen}). (Bottom) Variation in the energy threshold with Long Gate.}
 \label{caen_setting}
\end{figure}

In this paragraph, we describe the gate length setting of the CAEN DT5720 for the coincidence measurement. Figure $\ref{caen_setting}$ (Top) shows a schematic view of a signal waveform and gate timing. This board generates a trigger when the sampled ADC value exceeds the trigger threshold, and the waveform data are recorded as output data. The start and end of the recorded data can be set by Pre Gate and Long Gate values. In addition, the board outputs the maximum ADC value in the acquired waveform, which can be subsequently used to generate a pulse height spectrum. This board can output data by filtering the coincidence condition (a coincidence window width is discussed later). We set the trigger threshold to 40 $\mathrm{Ch}_{th}$ (107 keV equivalent for this case) unless otherwise noted. Here, the subscript "th" represents the trigger threshold channel of the CAEN DT 5720.

After measuring the rise of the waveform of the current waveform amplifier output with an oscilloscope, we set Pre Gate to 12 ns. Pertaining to the Long Gate, we identified the best gate width that allowed the energy threshold to obtain its lowest value. For this calculation, we used the Compton edge of $^{137}\mathrm{Cs}$ (see $\ref{sec:sample:appendix}$), and the results are shown in Figure $\ref{caen_setting}$ (Bottom). If the Long Gate width was longer, the noise pulse could be accidentally acquired, resulting in a significantly worse energy threshold. Therefore, we also set 100 ns as the best Long Gate. We will perform the measurements in the following sections with this Pre Gate and Long Gate.

\section{Energy Threshold When Using a Current Waveform Amplifier}\label{setu}
\begin{figure}[H]
 \centering
% \begin{subfigure}
   \centering
   \includegraphics[scale=0.7]{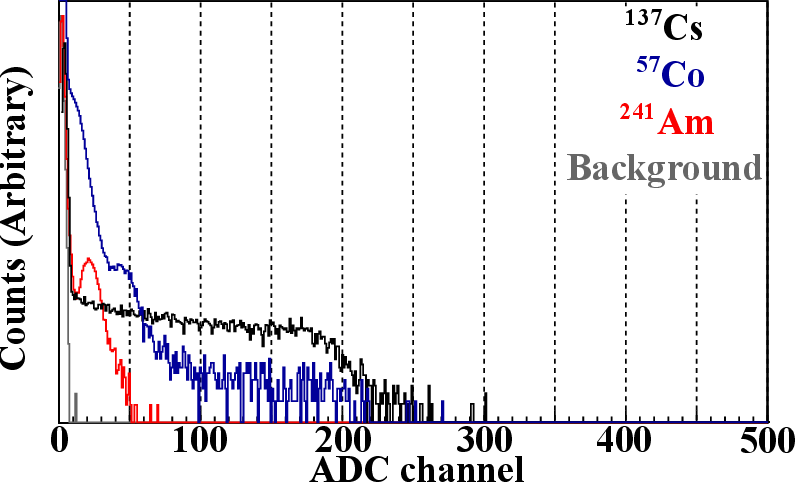}
   \label{3_8}
% \end{subfigure}\vfill
%  \begin{subfigure}
   \centering
   \includegraphics[scale=0.7]{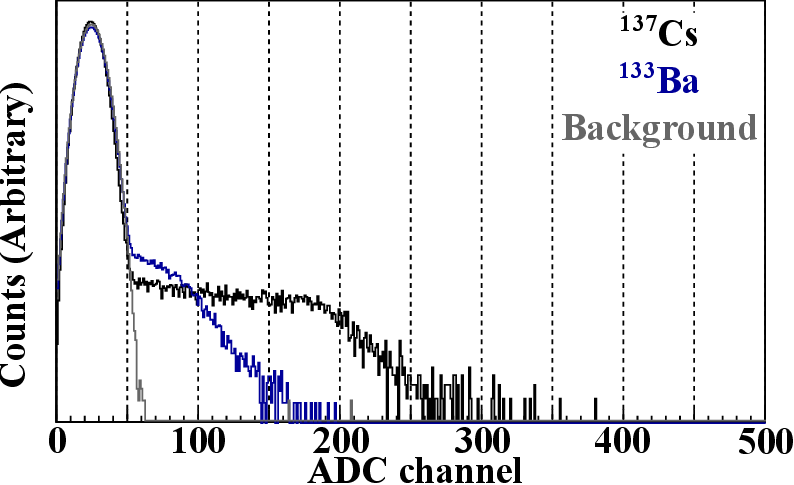}
   \label{3_7}
% \end{subfigure}\vfill
 \caption{Spectra obtained with a current waveform amplifier, setup in Figure $\ref{2_1}$. (Top) Normal SiPM, $^{137}\mathrm{Cs}$ (black), $^{57}\mathrm{Co}$ (blue), $^{241}\mathrm{Am}$ (red), and background (gray). (Bottom) Radiation-damaged SiPM, $^{137}\mathrm{Cs}$ (black), $^{133}\mathrm{Ba}$ (blue), and background (gray)}
 \label{normal_damaged}
\end{figure}

We initially compared the energy threshold obtained by one normal and one radiation-damaged SiPM, using the setup presented in Figure $\ref{2_1}$. Figure $\ref{normal_damaged}$ shows the pulse height spectra obtained by a normal SiPM and a radiation-damaged SiPM. With respect to the normal SiPM, the low noise levels allowed us to set the trigger threshold to 4 $\mathrm{Ch}_{th}$, which is 1/10 of that of the damaged SiPM. This is because a significantly greater number of noise events are bound to appear due to radiation damage when using a radiation-dominated SiPM.

As shown in Figure $\ref{normal_damaged}$ (Top), we determined a linear calibration curve from the data obtained after irradiating $^{57}\mathrm{Co}$ and $^{241}\mathrm{Am}$. In contrast, a linear calibration curve was determined by using the Compton edge for the $^{137}\mathrm{Cs}$ and $^{133}\mathrm{Ba}$ spectra, since the photo absorption peak was initially unclear (Figure $\ref{normal_damaged}$ [Bottom]). In this paper for determining the energy threshold, we measured with $^{137}\mathrm{Cs}$ source and found ADC channels where the signal is no longer buried in SiPM noise in the spectra. The calibration procedure is described in $\ref{sec:sample:appendix}$. We used the same $^{137}\mathrm{Cs}$ source at the same time for all measurements so that we minimized the fluctuation of the energy threshold values. We found that the energy thresholds for the normal and damaged SiPM were 23 and 132 keV, respectively, thereby confirming that the energy threshold is increased due to radiation damage.

\begin{figure}[H]
 \centering
% \begin{subfigure}
   \centering
   \includegraphics[scale=0.7]{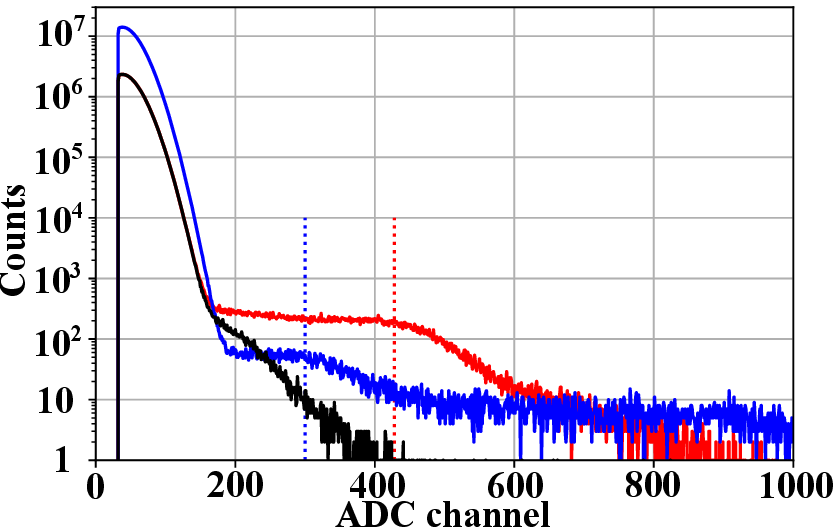}
   \label{6_3}
% \end{subfigure}\vfill
%  \begin{subfigure}
   \centering
   \includegraphics[scale=0.7]{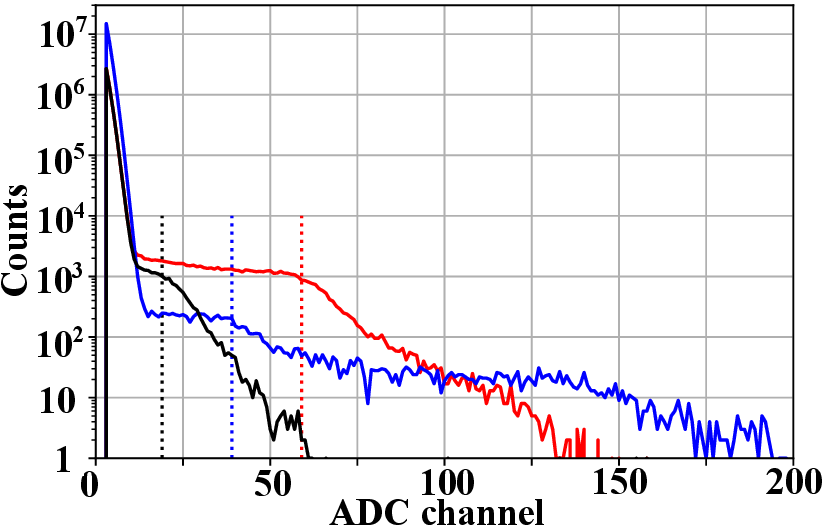}
   \label{6_4}
% \end{subfigure}\vfill
 \caption{Spectra of $^{137}\mathrm{Cs}$ (red),$^{22}\mathrm{Na}$ (blue), and $^{133}\mathrm{Ba}$ (black). (Top) Charge amplifier and shaping amplifier, upper setup in Figure $\ref{2_4}$. (Bottom) Current waveform amplifier, lower setup of the same figure. Dotted lines in these figures denote the location of Compton edges of $^{137}\mathrm{Cs}$, $^{22}\mathrm{Na}$, and $^{133}\mathrm{Ba}$, respectively; however the black line is not shown in the top figure because the energy threshold is almost the same as the Compton energy of $^{133}\mathrm{Ba}$.}
 \label{charge_current}
\end{figure}

The two figures in Figures $\ref{charge_current}$ illustrate the results of measurements that irradiated each source to the upper and lower setups with radiation-damaged SiPMs in Figure $\ref{2_4}$. These figures suggest that the energy threshold with a charge amplifier and a shaping amplifier was 198 keV, and 128 keV with a current waveform amplifier, and the latter is consistent with the value obtained with a radiation-damaged SiPM (Figure $\ref{normal_damaged}$ [Bottom]). Here, the current waveform amplifier just amplifies the current from SiPMs and converts it to a voltage signal whose waveform is the same as that of the current time profile. Consequently, the current waveform amplifier could lower the energy threshold to approximately 65$\%$ compared to charge-sensitive amplifier and shaping amplifier. We believe that because a plastic scintillator has a short decay time; and thus has a high signal to noise ratio and short integration time around the peak, the energy threshold can be lowered by capturing the peak of the shape pulse with the current waveform amplifier rather than integrating SiPM raw output with a charge-sensitive amplifier. Actually, the gain of the current waveform amplifier is lower than that of the charge-sensitive amplifier; however it also needs much lower consumption. Therefore, we consider it to be more suitable for long-term use in a satellite orbit. Regarding lower gain, we expect that it will not be a problem because we could detect the required signals. Furthermore, the current waveform amplifier provides worse energy resolution, but it is not a problem for anti-coincidence usage.

\section{Energy Threshold when Coinciding with two SiPMs}\label{setu2}
\subsection{Coincidence Measurement}\label{kou1}
\begin{figure}[H]
 \centering
% \begin{subfigure}
   \centering
   \includegraphics[scale=0.7]{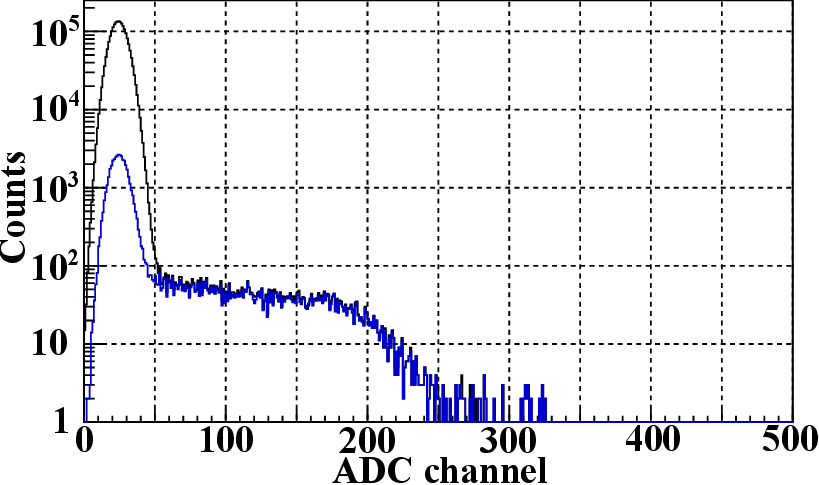}
   \label{5_1}
% \end{subfigure}\vfill
% \begin{subfigure}
   \centering
   \includegraphics[scale=0.7]{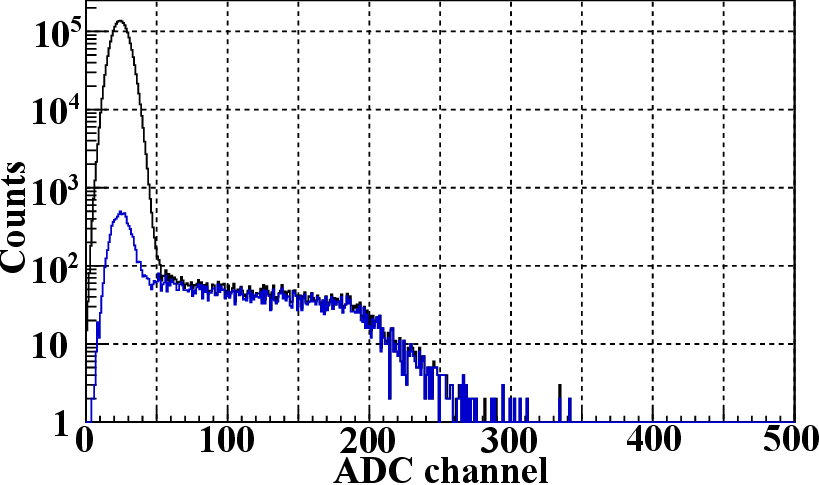}
   \label{5_4}
% \end{subfigure}\vfill
%  \begin{subfigure}
   \centering
   \includegraphics[scale=0.7]{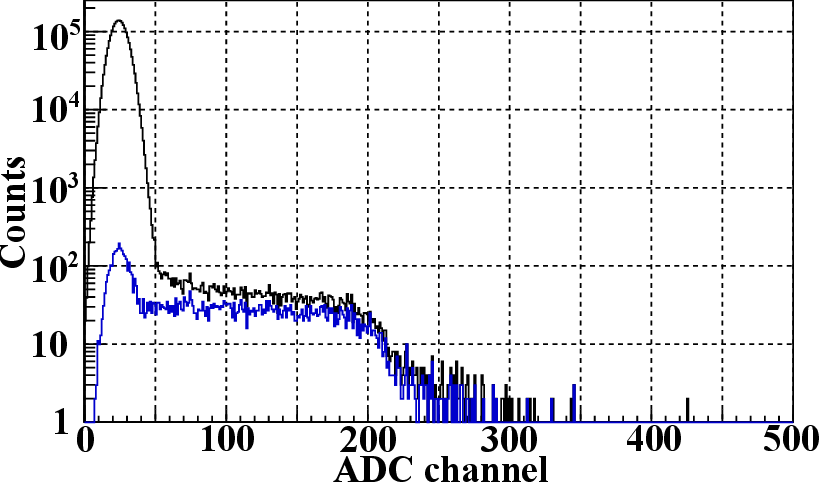}
   \label{5_2}
% \end{subfigure}\vfill
 \caption{Energy spectra of $^{137}\mathrm{Cs}$ with different coincidence widths of (Top) 40, (Middle) 8, and (Bottom) 4 ns, measured using setup in Figure $\ref{2_1}$. Blue and black lines represent the coincidence and noncoincidence spectra, respectively.}
 \label{three_plots}
\end{figure}

In this section, we describe the coincidence measurement using the setup shown in Figure $\ref{2_1}$. Since almost identical energy spectra were obtained for both SiPMs outputs, we only show the results for one of them. We performed this experiment by varying the coincidence width. Figure $\ref{three_plots}$ shows the spectra when setting the coincidence width to 40, 8, and 4 ns, and blue and black lines represent the coincidence and noncoincidence spectra of $^{137}\mathrm{Cs}$, respectively. As predicted in section $\ref{setu3}$ the energy threshold could be lowered. If we decide to significantly reduce the coincidence width, the energy threshold further decreases, but at the same time there is a significant signal loss (particularly when it comes to the low energy signal) as shown in Figure $\ref{three_plots}$ (Bottom).

\begin{figure}[H]
 \centering
 \includegraphics[scale=0.6]{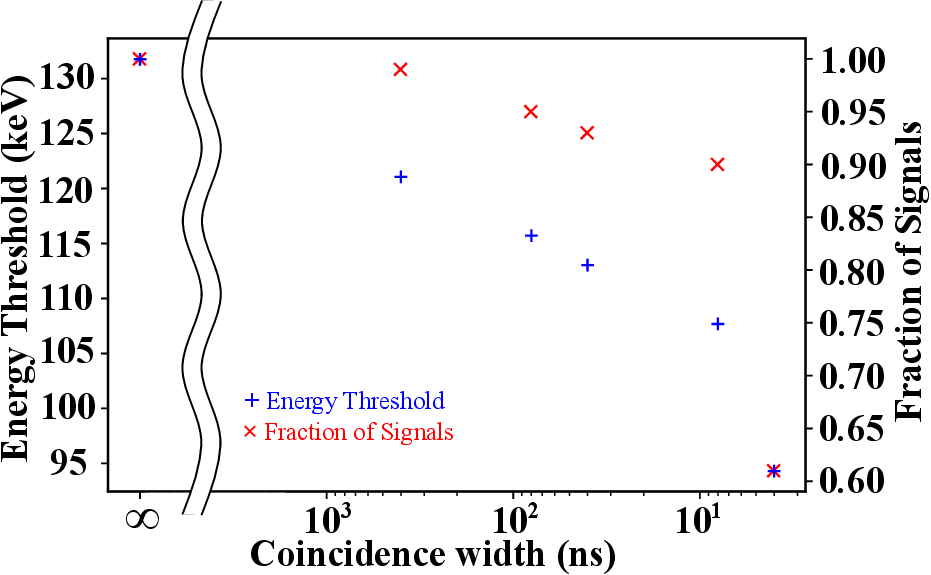}
 \caption{Variation of the energy threshold and fraction of detected signals with the coincidence width}
 \label{5_3}
\end{figure}

To identify the optimal coincidence width, we performed several measurements using different widths, and the relationship between the coincidence width and the energy threshold is summarized in Figure $\ref{5_3}$. The result obtained without any coincidence measurements in section $\ref{setu}$ is also shown at infinity. In this figure, we also calculated the fraction of the detected signals by comparing coincidence and noncoincidence measurements. As a signal, in this case we counted more than 54 ADC channels. When the coincidence width was set to 4 ns, the fraction of the acquired signal events was greatly reduced, and thus we determined that the optimal coincidence width using this setup was 8 ns, where the energy threshold was low and the fraction of the acquired signal events was approximately 90$\%$ (Figure $\ref{three_plots}$ [Middle]). This result demonstrates that the energy threshold can be lowered to 108 keV (approximately 82$\%$) using the coincidence measurement setup, compared to the case without coincidence. However, when inorganic scintillators with longer decay times than plastic scintillators are used, such as CsI, then a coincidence width of approximately 50-100 ns is required; and the energy threshold may increase compared to the present measurement. Therefore we consider that the proposed method is suitable for use of a plastic scintillator.

\begin{figure}[H]
 \centering
 \includegraphics[scale=0.7]{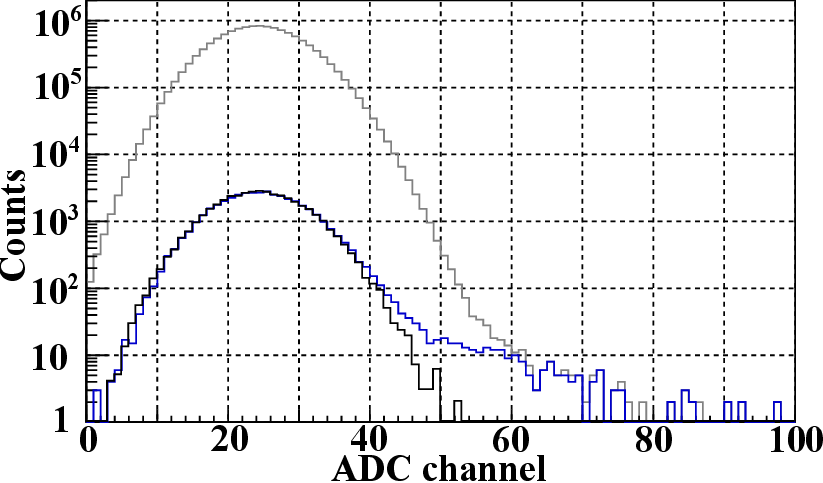}
 \caption{Spectra of $^{57}\mathrm{Co}$ (blue), background (black) with coincidence, and background without coincidence (gray), measured using setup in Figure $\ref{2_1}$.}
 \label{5_7}
\end{figure}

Since the energy threshold can be lowered using the previous method, the same setup could allow us to detect gamma-ray signals that are difficult to detect without coincidence measurements. Here, we measured the gamma-ray signal of a $^{57}\mathrm{Co}$ source and set the coincidence width to 8 ns as described above. As seen in Figure $\ref{5_7}$, we can find excess events around the ADC channel of 60-70, which are otherwise hidden under the noise events if we do not use coincidence measurements. Based on the energy calibration of Figure $\ref{normal_appendix}$ in $\ref{sec:sample:appendix}$, we confirmed that this excess corresponds to 122 keV from the gamma-ray signal of a $^{57}\mathrm{Co}$ source.

This coincidence method decreases the number of photons detected in our SiPM and thus we summed the signals from the two SiPMs attached to the same scintillator. In this case, the noise of SiPMs are also summed, making the signal-to-noise ratio worse. But this demerit is not important compared to eliminating the SiPM noise events.

\subsection{Pulse Heights for Each SiPM Signal}
\begin{figure}[H]
 \centering
% \begin{subfigure}
   \centering
   \includegraphics[scale=0.75]{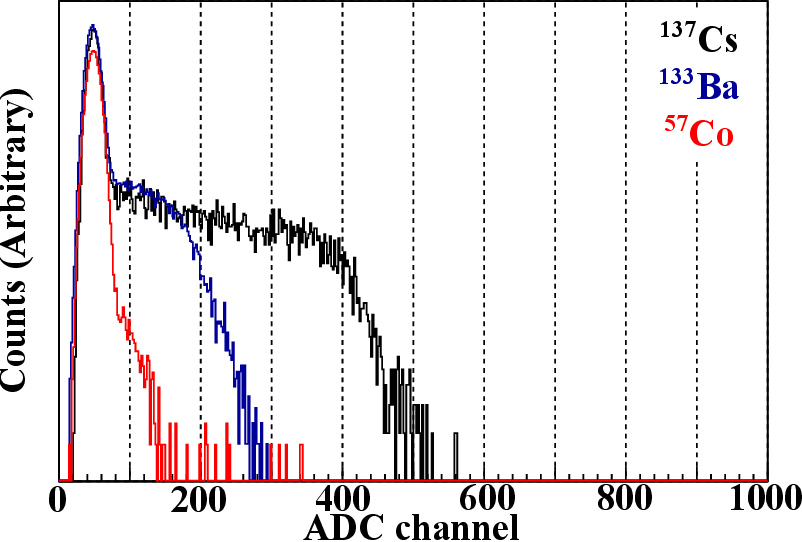}
   \label{5_6}
% \end{subfigure}\vfill
%  \begin{subfigure}
   \centering
   \includegraphics[scale=0.65]{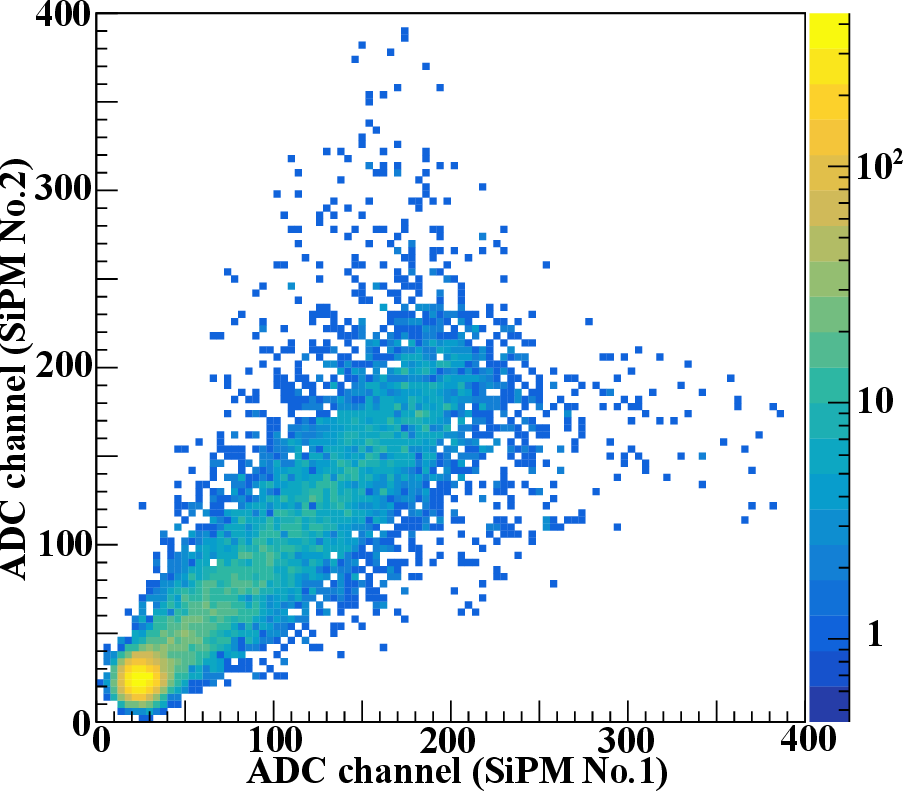}
   \label{5_5}
% \end{subfigure}\vfill
 \caption{(Top) Spectra obtained by summing up the pulse heights of the two SiPMs : $^{137}\mathrm{Cs}$ (black), $^{133}\mathrm{Ba}$ (blue), and $^{57}\mathrm{Co}$ (red). (Bottom) Spectra of the pulse heights obtained from the two SiPMs for $^{137}\mathrm{Cs}$, measured using in setup Figure $\ref{2_1}$.}
 \label{pulse_height}
\end{figure}

As mentioned above, we obtained almost the same spectra for the two SiPM outputs. However, to confirm this in practice we summed up the pulse heights of these two outputs for the same time event using the data shown in Figure $\ref{three_plots}$ (Middle) (Figure $\ref{pulse_height}$ [Top]). We obtained an energy threshold of 93 keV which is lower than that obtained from one SiPM spectrum (Figure $\ref{three_plots}$ [Middle]). We believe that discrepancy is due to the fact that the noise of one event is averaged out by summing up the noise from the two SiPMs.

In addition, we plotted the pulse heights of the two SiPMs for those coincident events on a two-dimensional histogram (Figure $\ref{pulse_height}$ [Bottom]). Although most events had similar pulse heights for both SiPMs, some events revealed that one SiPM had a much larger pulse height than the other one. We considered that these events occurred near the front of the SiPM that was attached to the scintillator\cite{uchida}.

\section{Measurement with Current Wave Amplifier at Low Temperature}\label{low2}
\begin{figure}[H]
 \centering
 \includegraphics[scale=0.7]{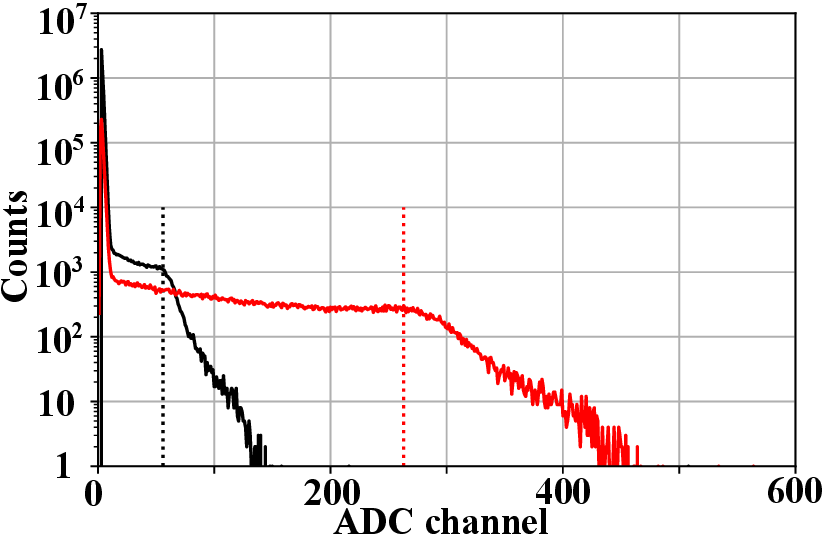}
 \caption{$^{137}\mathrm{Cs}$ spectra obtained using a current waveform amplifier with upper setup in Figure $\ref{2_4}$ at +20 $^\circ$C (black) and -20 $^\circ$C (red), respectively. Dotted lines denote the location of $^{137}\mathrm{Cs}$ Compton edges (477 keV).}
 \label{low}
\end{figure}

The measurement was performed at a low temperature of -20 $^\circ$C; using the current waveform amplifier. Here the bias voltage of the SiPM was set at 54.75 V ($\sim$$V_{br}+5$V) to maintain its value same as that measured at room temperature. Figure $\ref{low}$ illustrates a comparison of $^{137}\mathrm{Cs}$ spectra obtained at room temperature and at -20 $^\circ$C. These measurements indicate that the energy threshold was 43 keV at -20 $^\circ$C. This is approximately 34$\%$ of the threshold energy observed at the room temperature. The temperature dependence of the light yield of plastic scintillator is at most 4$\%$\cite{yield}, implying that this threshold change is mostly because of a reduction in the SiPM noise. The dark current of this SiPM is $\sim$0.3 and $\sim$0.1 mA at room temperature and -20 $^\circ$C, respectively. Therefore, the dark current also reduces by a factor of 3, almost consistent with the reduction in the threshold energy.

\section{Summary}\label{summary}
In this paper, we attempted to lower the energy threshold of a plastic scintillator and radiation-damaged SiPMs using three different methods. More specifically, the energy threshold was lowered to approximately 65$\%$ (from 198 keV to 128 keV) using the current waveform amplifier instead of the charge-sensitive amplifier and shaping amplifier, approximately 70$\%$ (from 132 keV to 93 keV) by only measuring coincident events and summing up the pulse heights of the two coincident SiPMs compared to the use of only one scintillator, and it reduced to approximately 34$\%$ (from 128 keV to 43 keV) when the measurement was performed at a low temperature of -20 $^\circ$C compared to that performed at 20 $^\circ$C. Table $\ref{7_2}$ summarizes the energy thresholds found in the present study. Pertaining to the coincidence measurements, we conclude that the optimal coincidence width for this plastic scintillator was 8 ns due to its short decay time.

\begin{table}[htbp]
\centering
\begin{tabular}{c|c} \hline
Measurement Condition & Threshold ($\mathrm{keV}$)\\ \hline
Damaged (Charge Amplifier + Shaping Amplifier, MCA) & 198 \\ 
Damaged (Current Waveform Amplifier, MCA) & 128 \\  
Normal (Current Waveform Amplifier, CAEN) & 23 \\
Damaged (Current Waveform Amplifier, CAEN) & 132 \\ 
Damaged (Current Waveform Amplifier, Coincidence, CAEN) & 108  \\
Damaged (Summing up Coincidence Spectra, CAEN) & 93  \\
Damaged (Current Waveform Amplifier, MCA, -20 $^\circ$C) & 43 \\ \hline
\end{tabular}
\caption{Summary of the energy thresholds found in the present study. Normal reflects the normal SiPM and Damaged reflects the damaged-SiPM.}
\label{7_2}
\end{table}

We can expect that the process of summing up and averaging the signals would allow us to lower the energy threshold by canceling out and relatively reducing the noise. Finally, we conclude that the combination of these three methods could lower the energy threshold to approximately 15$\%$. Therefore, we can effectively maintain the threshold of the detectors as low as possible using the appropriate setup.

\appendix

\section{How to Determine the Location of Compton Edges}
\label{sec:sample:appendix}

\begin{figure}[H]
 \centering
% \begin{subfigure}
   \centering
   \includegraphics[scale=0.7]{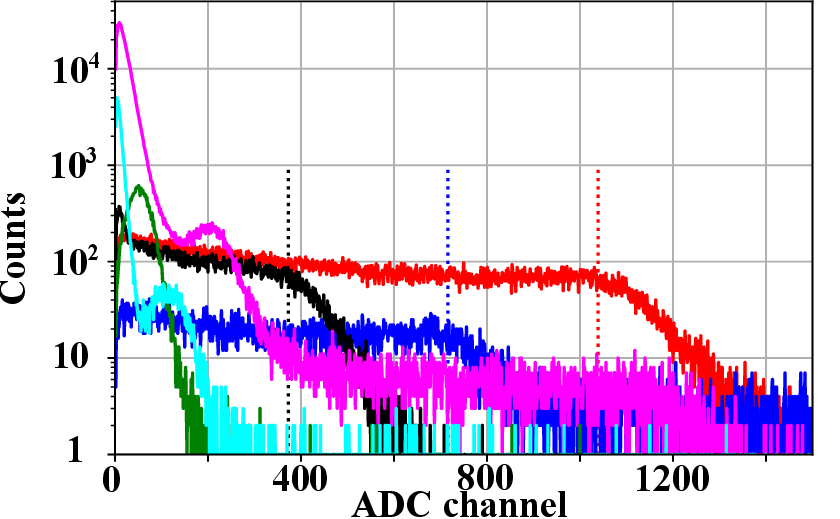}
   \label{a_1}
% \end{subfigure}\vfill
%  \begin{subfigure}
   \centering
   \includegraphics[scale=0.7]{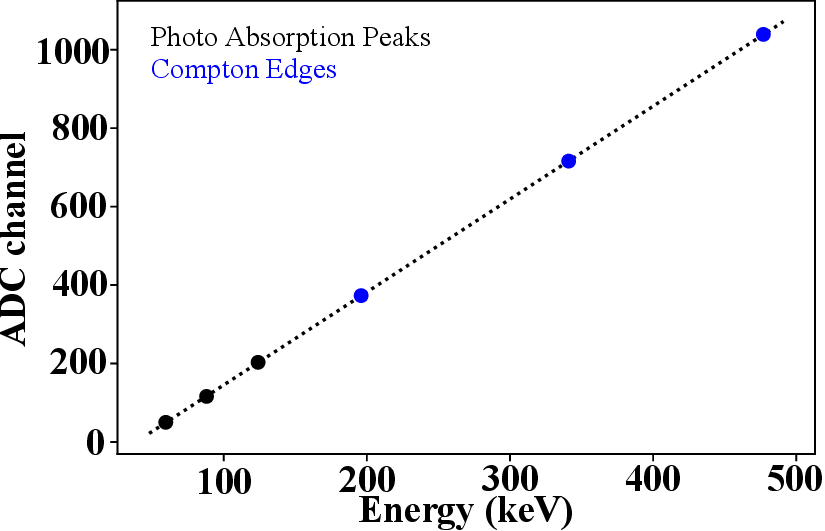}
   \label{a_2}
% \end{subfigure}\vfill
 \caption{(Top) Spectra of $^{137}\mathrm{Cs}$ (red), $^{22}\mathrm{Na}$ (blue), $^{133}\mathrm{Ba}$ (black), $^{57}\mathrm{Co}$ (magenta), $^{241}\mathrm{Am}$ (green), and $^{109}\mathrm{Cd}$ (cyan) with a normal SiPM and a current waveform amplifier with upper setup in Figure $\ref{2_4}$. Dotted lines denote the location of Compton edges of $^{137}\mathrm{Cs}$, $^{22}\mathrm{Na}$, and $^{133}\mathrm{Ba}$, respectively. (Bottom) Linear calibration curve obtained from this top figure.}
 \label{normal_appendix}
\end{figure}

Here, we describe the method with which we determined the location of the Compton edges in the spectra of the radiation-damaged SiPMs. First, we made a linear calibration curve of the normal SiPM with energy sources to allow us to resolve the exact photo absorption peaks ($^{57}\mathrm{Co}$, $^{109}\mathrm{Cd}$, and $^{241}\mathrm{Am}$). Based on this linear calibration curve, we determined the location of the Compton edges of $^{137}\mathrm{Cs}$, $^{22}\mathrm{Na}$, and $^{133}\mathrm{Ba}$, as shown in Figure $\ref{normal_appendix}$ (dotted lines represent their locations, respectively). In this measurement, we attached only one SiPM to the scintillator to increase the light intensity and applied voltage of 53.2 V ($V_{br}+2 V$) to the setup to avoid saturation.

When we decide to use radiation-damaged SiPM, then the low-energy photo absorption peaks cannot be clearly detected and the energy resolution of the Compton edges is bound to become worse. Consequently, we determined the location of Compton edges by convolving the spectra of Figure $\ref{normal_appendix}$ (Top) with a Gaussian curve with the energy resolution of the radiation-damaged SiPMs and superimposed on the spectra acquired with the radiation-damaged SiPM. The error caused by this method is estimated to be almost 5 ADC channels, which is approximately 2$\%$-3$\%$ in energy conversion.

\section{Light Yield of Scintillator}
\begin{figure}[H]
 \centering
 \includegraphics[scale=0.7]{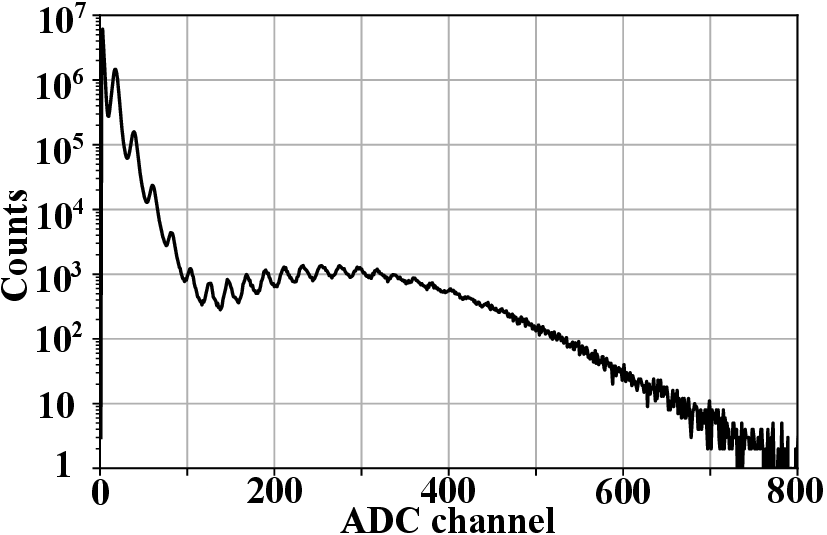}
 \caption{$^{109}\mathrm{Cd}$ spectra obtained using a charge-sensitive amplifier and a shaping amplifier at -20 $^\circ$C with lower setup in Figure $\ref{2_4}$.}
 \label{6_2}
\end{figure}

We measured the light yield of the scintillator used in this study. We performed a measurement using the charge-sensitive amplifier and the shaping amplifier as shown in Figure $\ref{2_4}$, and utilized the normal SiPM at -20 $^\circ$C. The spectrum obtained by irradiating $^{109}\mathrm{Cd}$ is shown in Figure $\ref{6_2}$. Counting the peaks for each photon, we found that the photo absorption peak at 22 keV corresponds to 14 photons. This means that the light yield of this plastic scintillator is two-third of photons per keV and this value is smaller than 10 photons per keV in \cite{ej200}. However, this result is appropriate considering the quantum efficiency and effective photosensitive area size of this SiPM, and the fact we attached SiPMs to both sides of the scintillator.

\section*{Acknowledgements}
This work was supported by JSPS KAKENHI, Japan Grant Numbers JP21KK0051 and JP19H01908.

%% The Appendices part is started with the command \appendix;
%% appendix sections are then done as normal sections
%\appendix

%\section{Sample Appendix Section}
%\label{sec:sample:appendix}

%% If you have bibdatabase file and want bibtex to generate the
%% bibitems, please use
%%
\bibliographystyle{elsarticle-num.bst} 
\bibliography{cas-refs.bib}

\begin{thebibliography}{10}
\expandafter\ifx\csname url\endcsname\relax
  \def\url#1{\texttt{#1}}\fi
\expandafter\ifx\csname urlprefix\endcsname\relax\def\urlprefix{URL }\fi
\expandafter\ifx\csname href\endcsname\relax
  \def\href#1#2{#2} \def\path#1{#1}\fi

\bibitem{bgo}
Regarding {BGO} scintillator, available at, \url{https://www.isas.jaxa.jp/missions/spacecraft/past/hitomi.html}.

\bibitem{rad}
X.~Zheng, H.~Gao, J.~Wen, M.~Zeng, X.~Pan, D.~Xu, Y.~Liu, Y.~Zhang, H.~Peng, Y.~Jiang, X.~Long, D.~Lu, D.~Yang, H.~Feng, Z.~Zeng, J.~Cang, Y.~Tian, In-orbit radiation damage characterization of sipms in the grid-02 cubesat detector, Nuclear Instruments and Methods in Physics Research Section A: Accelerators, Spectrometers, Detectors and Associated Equipment 1044 (2022) 167510.

\bibitem{camelot}
J.~^^c5^^98^^c3^^adpa, G.~Galg^^c3^^b3czi, N.~Werner, A.~P^^c3^^a1l, M.~Ohno, L.~M^^c3^^a9sz^^c3^^a1ros, T.~Mizuno, N.~Tarcai, K.~Torigoe, N.~Uchida, Y.~Fukazawa, H.~Takahashi, K.~Nakazawa, N.~Hirade, K.~Hirose, S.~Hisadomi, T.~Enoto, H.~Odaka, Y.~Ichinohe, Z.~Frei, L.~Kiss, Estimation of the detected background by the future gamma ray transient mission camelot, Astronomische Nachrichten 340 (2019) 666--673.

\bibitem{hirade}
N.~Hirade, H.~Takahashi, N.~Uchida, M.~Ohno, K.~Torigoe, Y.~Fukazawa, T.~Mizuno, H.~Matake, K.~Hirose, S.~Hisadomi, K.~Nakazawa, K.~Yamaoka, N.~Werner, J.~^^c5^^98^^c3^^adpa, S.~Hatori, K.~Kume, S.~Mizushima, Annealing of proton radiation damages in si-pm at room temperature, Nuclear Instruments and Methods in Physics Research Section A: Accelerators, Spectrometers, Detectors and Associated Equipment 986 (2021) 164673.

\bibitem{SiPM}
{K.K. Hamamatsu Photonics datasheet}, available at, \url{https://www.hamamatsu.com/content/dam/hamamatsu-photonics/sites/documents/99_SALES_LIBRARY/ssd/mppc_kapd9008e.pdf}.

\bibitem{ej200}
Regarding {EJ-200}, available at, \url{https://eljentechnology.com/products/plastic-scintillators/ej-200-ej-204-ej-208-ej-212}.

\bibitem{takahashi}
H.~Takahashi, N.~Hirade, N.~Uchida, K.~Hirose, T.~Mizuno, Y.~Fukazawa, K.~Yamaoka, H.~Tajima, M.~Ohno, Silicon photomultiplier (si-pm) comparisons for low-energy gamma ray readouts with bgo and csi (tl) scintillators, Nuclear Instruments and Methods in Physics Research Section A: Accelerators, Spectrometers, Detectors and Associated Equipment 989 (2021) 164945.

\bibitem{torigoe}
K.~Torigoe, Y.~Fukazawa, G.~Galg^^c3^^b3czi, T.~Mizuno, K.~Nakazawa, M.~Ohno, A.~P^^c3^^a1l, H.~Takahashi, K.~Tanaka, N.~Tarcai, N.~Uchida, N.~Werner, T.~Enoto, Z.~Frei, Y.~Ichinohe, L.~Kiss, H.~Odaka, J.~^^c5^^98^^c3^^adpa, Z.~V^^c3^^a1rhegyi, Performance study of a large csi(tl) scintillator with an mppc readout for nanosatellites used to localize gamma-ray bursts, Nuclear Instruments and Methods in Physics Research Section A: Accelerators, Spectrometers, Detectors and Associated Equipment 924 (2019) 316--320.

\bibitem{mca}
{MCA 8000D} manual, available at, \url{https://www.amptek.com/-/media/ametekamptek/documents/resources/products/user-manuals/mca8000d-user-manual-b1.pdf?la=en&revision=75b93881-a2e4-4d92-9d84-0d67b5d34614}.

\bibitem{caen}
{CAEN DT5720} manual, available at, \url{https://www.caen.it/products/dt5720/}.

\bibitem{wakasa}
S.~Hatori, T.~Kurita, Y.~Hayashi, M.~Yamada, H.~Yamada, J.~Mori, H.~Hamachi, S.~Kimura, T.~Shimoda, M.~Hiroto, T.~Hashimoto, M.~Shimada, H.~Yamamoto, N.~Ohtani, K.~Yasuda, R.~Ishigami, M.~Sasase, Y.~Ito, M.~Hatashita, K.~Takagi, K.~Kume, S.~Fukuda, N.~Yokohama, G.~Kagiya, S.~Fukumoto, M.~Kondo, Developments and applications of accelerator system at the wakasa wan energy research center, Nuclear Instruments and Methods in Physics Research Section B: Beam Interactions with Materials and Atoms 241 (2005) 862--869.

\bibitem{uchida}
N.~Uchida, H.~Takahashi, M.~Ohno, T.~Mizuno, Y.~Fukazawa, M.~Yoshino, K.~Kamada, Y.~Yokota, A.~Yoshikawa, Attenuation characteristics of a $\mathrm{Ce:Gd_{3}Al_{2}Ga_{3}O_{12}}$ scintillator, Nuclear Instruments and Methods in Physics Research Section A: Accelerators, Spectrometers, Detectors and Associated Equipment 986 (2021) 164725.

\bibitem{yield}
L.~Peralta, Temperature dependence of plastic scintillators, Nuclear Instruments and Methods in Physics Research Section A: Accelerators, Spectrometers, Detectors and Associated Equipment 883 (2018) 20--23.

\end{thebibliography}

%\begin{thebibliography}{99}

%% else use the following coding to input the bibitems directly in the
%% TeX file.

% \begin{thebibliography}{00}

% %% \bibitem{label}
% %% Text of bibliographic item

% \bibitem{}

\end{document}